\documentclass[aps,preprint,showpacs,floatfix]{revtex4}
\usepackage{graphicx,bm,amsmath,dcolumn}
\usepackage{amsmath,amssymb,amsthm,graphicx,epsfig}

\newcommand{\gae}
{\,\hbox{\lower0.5ex\hbox{$\sim$}\llap{\raise0.5ex\hbox{$>$}}}\,}
\newcommand{\lae}
{\,\hbox{\lower0.5ex\hbox{$\sim$}\llap{\raise0.5ex\hbox{$<$}}}\,}
\newcommand{\be}{\begin{equation}}
\newcommand{\ee}{\end{equation}}
\newcommand{\bea}{\begin{eqnarray}}
\newcommand{\eea}{\end{eqnarray}}
\newcommand{\bdm}{\begin{displaymath}}
\newcommand{\edm}{\end{displaymath}}

\newcommand{\rmc}{{\rm c}}

\begin{document}
\title{The three-state Potts model on the centered triangular lattice}
\author{Zhe Fu}
\affiliation{College of Physics and Electronic Engineering,
Xinxiang University, Xinxiang 453003, China}
\author{Wenan Guo}
\email{waguo@bnu.edu.cn}
\affiliation{Physics Department, Beijing Normal University,
Beijing 100875, China}     
\affiliation{Beijing Computational Science Research Center, Beijing 100193,
China}
\author{Henk W.~J. Bl\"ote}
\email{henk@lorentz.leidenuniv.nl}
\affiliation{Instituut Lorentz, Leiden University, P.O. Box 9506,
2300 RA Leiden, The Netherlands}
\begin{abstract}
We study phase transitions of the Potts model
on the centered-triangular lattice with two types of couplings,
namely $K$ between neighboring triangular sites, and $J$ between the 
centered and the triangular sites.
Results are obtained by means of a finite-size analysis based on numerical
transfer-matrix calculations and Monte Carlo simulations.
Our investigation covers the whole $(K,J)$ phase diagram, but we find that
most of the interesting physics  applies to the antiferromagnetic case $K<0$,
where the model is geometrically frustrated. 
In particular, we find that there are, for all finite $J$, two transitions
when $K$ is varied. Their critical properties are explored.
In the limits $J\to \pm \infty$ we find algebraic phases
with infinite-order transitions to the ferromagnetic phase.

\end{abstract}
\maketitle

\section{Introduction}
The Potts model \cite{Potts} is defined in terms of $q$-state lattice
variables, also called spins, $\sigma_i=1,2,\ldots,q$, where $i$ stands
for the lattice site of the variable.  Neighboring spins interact only if
they are equal. Since its introduction, the model has played a significant
role in statistical physics \cite{esmsm,CFT}, and in applications to
various condensed-matter systems \cite{Wurev}.

Originally, most studies of the Potts model focused on ferromagnetic
interactions, and for that case the critical properties and phase diagram
are well known. However, more recently also the antiferromagnetic(AF)
Potts model has received considerable attention, because of its rich
and lattice-dependent behavior. For instance, the behavior of the AF
$q=3$ Potts model on several lattices appears to be quite different. 
The model displays a weak first-order transition at a nonzero
temperature on the triangular lattice \cite{Adler},
an ordinary finite-temperature critical point on the diced lattice
\cite{dicedRK}, and on the honeycomb lattice it is disordered at all
non-negative temperatures \cite{honeycombJS}. 
On the square lattice, it is critical at zero temperature, and disordered
at positive temperatures \cite{BH, Kolafa, denNijs,SS}.
On a set of planar lattices called quadrangulations
the model either has a zero-temperature critical point, or it has
three ordered coexisting phases, dependent on whether or not the
quadrangulation is self-dual \cite{jplv}.
In view of this lattice-dependent behavior, AF Potts models have to be
investigated case by case.

From another point of view, AF Potts models on many regular lattices
have an interesting feature: there exists a lattice-dependent critical
value $q_\rmc$ of $q$ beyond which there is no transition.
The generalization of the Potts model to the random-cluster model \cite{KF},
in which $q$ is a continuous variable, enables the determination of
$q_\rmc$ even if it is not an integer.
For example, $q_\rmc=\frac{1}{2}(3+\sqrt{5})$ for the honeycomb lattice
was determined \cite{WG} by examining the known critical frontiers in
the light of AF interactions.
However, Huang {\it et al} \cite{Huang} have discovered a set of
lattices on which the AF Potts model does not have such a $q_\rmc$.
Furthermore, some AF Potts models on irregular lattices, in which
the number of sites is different for different sublattices, display
entropy-driven transitions at a finite temperature to partially ordered
phases at a value of $q$ larger than the $q_\rmc$ that one would naively
expect \cite{dicedRK,kotecky1,Chen,Deng,Huang}.

\begin{figure}[bthp]
\begin{center}
\leavevmode
\includegraphics[width=8.4cm,angle=0]{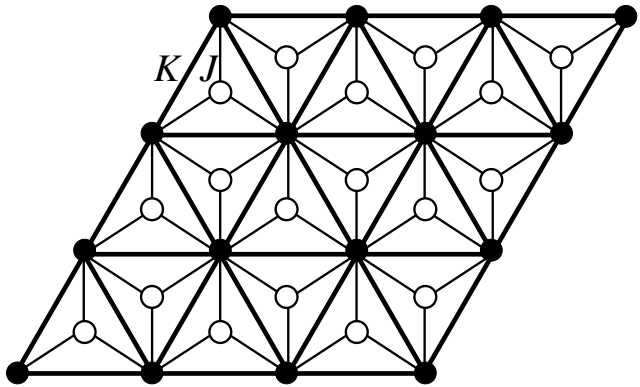}
\end{center}
\caption{The centered triangular lattice. The triangular vertices
($\bullet$) as well as the centered vertices ($\circ$) are occupied by
three-state Potts spins. Neighboring spins on the triangular sites are
coupled with strength $K$, and the centered sites are coupled with
strength $J$ to their triangular neighbors.}
\label{ctril}
\end{figure}

The present work considers the case of the $q=3$ model on the centered
triangular lattice, also known as the asanoha or hemp-leaf
lattice \cite{Syozi}, which is sketched in Fig.~\ref{ctril}. \\

The interactions are specified by the reduced Hamiltonian
\begin{equation}
{ H}/k_{\rm B}T = -  K \sum_{<i,j>} \delta_{\sigma_i \sigma_j}
                  -  J \sum_{[k,l]} \delta_{\sigma_k \sigma_l}
\end{equation}
where the sum on ${<i,j>}$ runs over all bonds connecting nearest-neighbor
spins on the triangular sites, and the sum on ${[k,l]}$ runs over all bonds
between the centered spins and their three triangular neighbor spins.
The corresponding Potts couplings are denoted by $K$ and $J$.
In the case $J=0$ the model reduces to the $q=3$ Potts model on the
triangular lattice. For $K=0$ the model reduces to the $q=3$ Potts model
on the diced lattice.

\section{Algorithms and tests}

A transfer-matrix algorithm using the $q=3$ spin representation
was employed for the calculation of the free energy densities
and magnetic correlation lengths for finite sizes up to $L=18$.
The spin systems studied were wrapped on $L \times \infty$ cylinders,
with periodic boundary conditions in the finite direction, using a
length unit equal to the triangular edges.
The transfer-matrix algorithm is applicable for all $J$ and $K$.
It does, in most cases, allow rather accurate determinations of
phase transitions and some universal parameters.
In those cases where we did not require very precise results, for 
instance for the global determination of phase boundaries, we also
applied a Metropolis-type Monte Carlo algorithm.

\subsection{Miscellaneous results of the transfer-matrix algorithm}
In the case $J=0$ the model reduces to the $q=3$ Potts model on the
triangular lattice. We first consider the ferromagnetic case $K>0$,
and required that the magnetic correlation lengths $\xi(K,L)$ satisfies
Cardy's asymptotic relation \cite{Cardyxi} $L/\xi(K,L)\simeq 2\pi X_h$,
where $X_h=2/15$ is the exactly known \cite{CG} magnetic dimension.
We solved $K$ for each value of $2<L\leq 18$, and thus obtained a 
series of estimates of the critical point. Extrapolation by finite-size
scaling \cite{FSS}, using correction
exponents $y_{\rm irr}-y_t=-2$  \cite{CG} and $-4$, led to a best
estimate $K_{\rm c,triangular}=0.630944725 (5)$. This value is close to the
exactly known critical point $\ln[2 \cos(\pi/9)]$ \cite{Baxtrcp,Wurev},
thus providing a consistency check.
For $K=0$ the model reduces to the $q=3$ Potts model on the diced
lattice. A similar analysis yielded finite-size estimates of its 
ferromagnetic critical point $J_{\rm c,diced}$ in the range $2<L\leq 18$. 
Extrapolation led to a best
estimate $J_{\rm c,diced}=0.955032665 (5)$. This value is close to an 
unpublished transfer-matrix result $J_{\rm c,diced}=0.9550325 (23)$ as 
quoted by Wu and Guo \cite{WG}. Our result for the diced lattice also
yields, by duality, the critical coupling of the $q=3$ Potts model on
the  kagome lattice as $K_{\rm c,kagome}=1.056560222 (5)$.
This is in agreement with 1.05656027 (7) as obtained by Jacobsen and
Scullard \cite{JS}.
Furthermore we performed a similar analysis for the antiferromagnetic
$q=3$ Potts model on the diced lattice, from which we estimate
% -3.9407892 (10)
$J_{\rm c,diced~AF}=-1.9703946 (5)$.

\section{Phase diagram in the $(K,J)$ plane}
%We have explored the global phase diagram using Monte Carlo simulations.
One can distinguish three different regions, according to the relative
magnitudes of the weights $W_{\rm 111}$ of a triangle with three equal spins,
$W_{\rm 112}$ for only two equal spins, and $W_{\rm 123}$ for three
different spins.
Since each $K$ coupling is shared between two triangles, only one half of it
is included in these weights. Furthermore the centered spins are summed out,
so that the weights  depend only on the triangular spins, while they still
include the effect of $J$: 
\begin{eqnarray}
W_{\rm 111} &=& \exp(3K/2+3J)+2\exp(3K/2)  \nonumber \\
W_{\rm 112} &=& \exp(K/2+2J)+\exp(K/2+J)+\exp(K/2) \nonumber  \\
W_{\rm 123} &=& 3\exp(J) \, .  \nonumber
\label{C2fs}
\end{eqnarray}
For $J>>0$ we have $\ln W_{\rm 111}\simeq 3K/2+3J$. For $J<<0$ the centered
spins will assume a state different from their triangular neighbors, so that
$\ln W_{\rm 111}\simeq 3K/2$. For $K$ sufficiently large negative, the weight
$W_{\rm 123}$ will dominate, and frustration of the centered spins will lead to
$\ln W_{\rm 123}\simeq J$. One also expects an intermediate region dominated
by triangles having two equal spins, with $\ln W_{\rm 112}\simeq K/2+2J$ for
$J>>0$ and $K/2$ for $J<<0$.

\begin{figure}[bthp]
\begin{center}
\leavevmode
\includegraphics[width=9.6cm,angle=0]{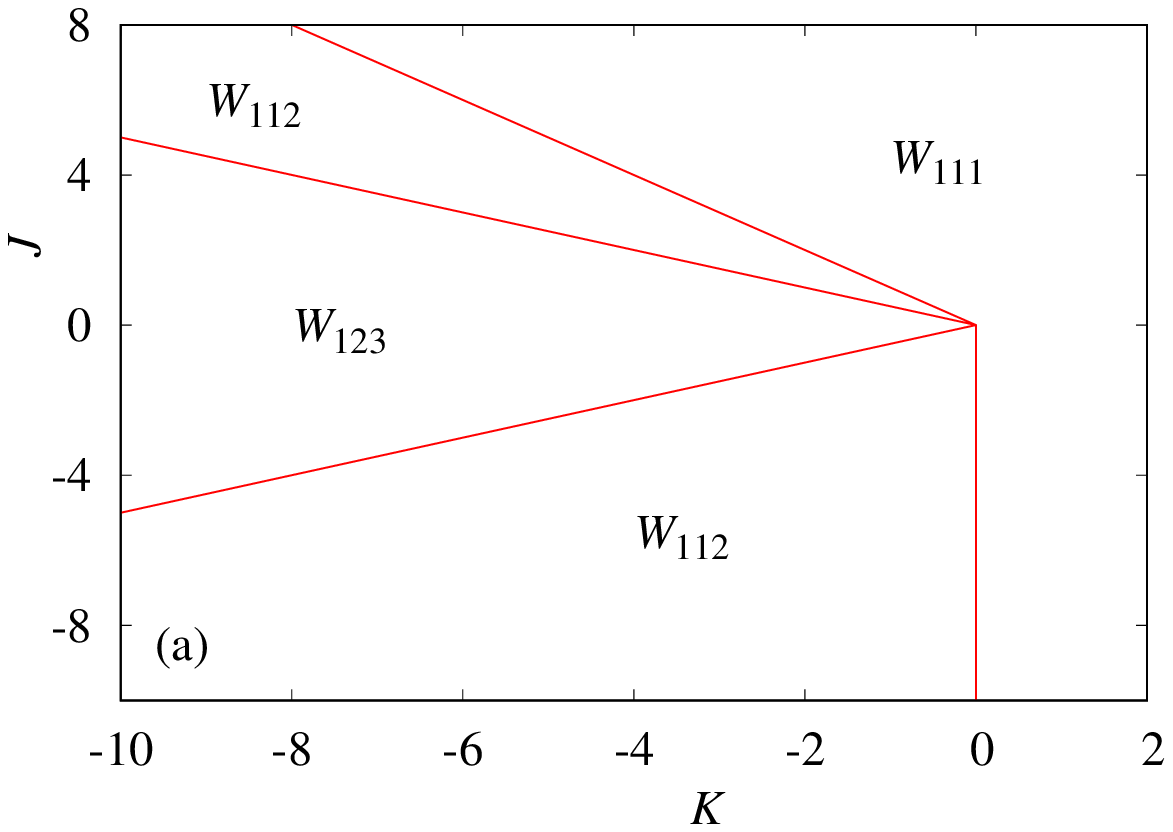}\\
\includegraphics[width=9.6cm,angle=0]{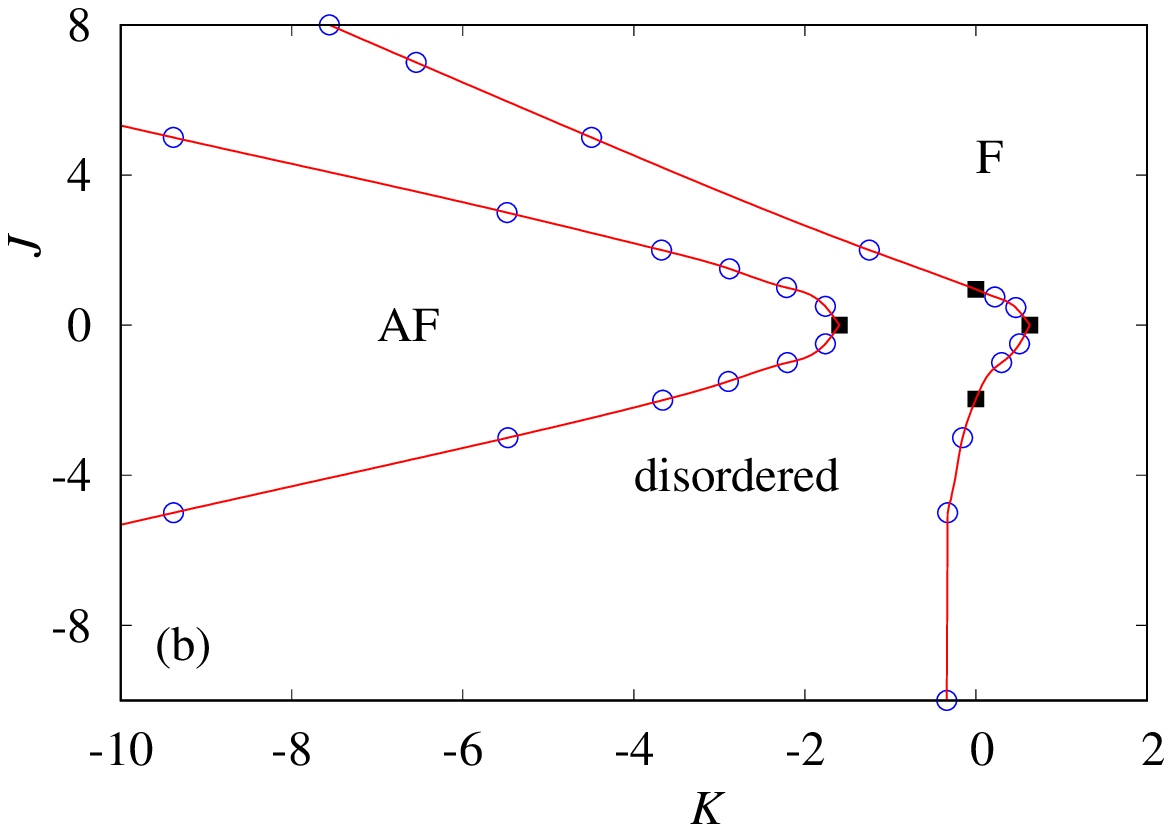}
\end{center}
\caption{Phase diagram of the centered triangular $q=3$ Potts model. 
Figure (a) shows the division of the ($K,J$) diagram according to the
dominance of the leading terms in the weights $W_{\rm 111}$,
$W_{\rm 112}$, and $W_{\rm 123}$ described in the text.
Figure (b) displays the numerical results for the
phase boundaries, obtained by Monte Carlo simulations, except those
shown as black squares ({\tiny $\blacksquare$}) which are accurate
results obtained by other methods, as mentioned in the text.
For all finite $J$, one observes three different phases: an 
antiferromagnetic (AF) one, an intermediate (disordered) one, and a
ferromagnetic (F) one.
}
\label{phdia}
\end{figure}

The phase boundaries are approximately located where the weights of
two neighboring phases become equal.
Thus we expect the following phases, shown in Fig.~\ref{phdia}(a):
\begin{enumerate}
\item
the ferromagnetic region, dominated by the weight $W_{\rm 111}$.  For $J>0$ it
is located at $K \gae -J$, and for $J<0$   at $K \gae 0$.
\item
the intermediate region, dominated by the weight $W_{\rm 112}$.  For $J>0$ it
is located at $-2J \lae K \lae -J$, and for $J<0$ at $2J \lae K \lae 0$.
\item
the antiferromagnetic region, dominated by the weight $W_{\rm 123}$. 
For both signs of $J$ it is located at $K \lae -2|J|$.
\end{enumerate} 
Monte Carlo, exact, and transfer-matrix results, shown in Fig.~\ref{phdia}(b),
confirm this expectation.
It appears that the intermediate phase is disordered, at least as long 
as $|J|$ is not too large. Partial order appears for large $|J|$, whose
nature will be explored in the following subsections.

Ferromagnetic $q=3$ universality applies naturally to the transition line
between regions 1 and 2.
As for the antiferromagnetic transition line, the triangular model
at $J=0$ was found to undergo a weak first-order transition,
see for instance Adler {\it et al.}~\cite{Adler} and references therein. 
This transition is located near $K=-1.594482 (8)$ \cite{Wea}.

\subsection{Mapping on the honeycomb O(2) loop model}
\label{mapon}
In the special case  of the limits
\begin{equation}
|J| \to \infty\, ,~~~~ K>> -2|J|, 
\label{Jinf}
\end{equation}
the spin model becomes equivalent with the nonintersecting O(2) loop
model on the honeycomb lattice. That model displays a range where the
magnetic correlation function decays algebraically \cite{N82}.
This proves that the spin model must reach a critical state at the
corresponding parameters.

\begin{figure}[bthp]
\begin{center}
\leavevmode
\includegraphics[width=7.2cm,angle=180]{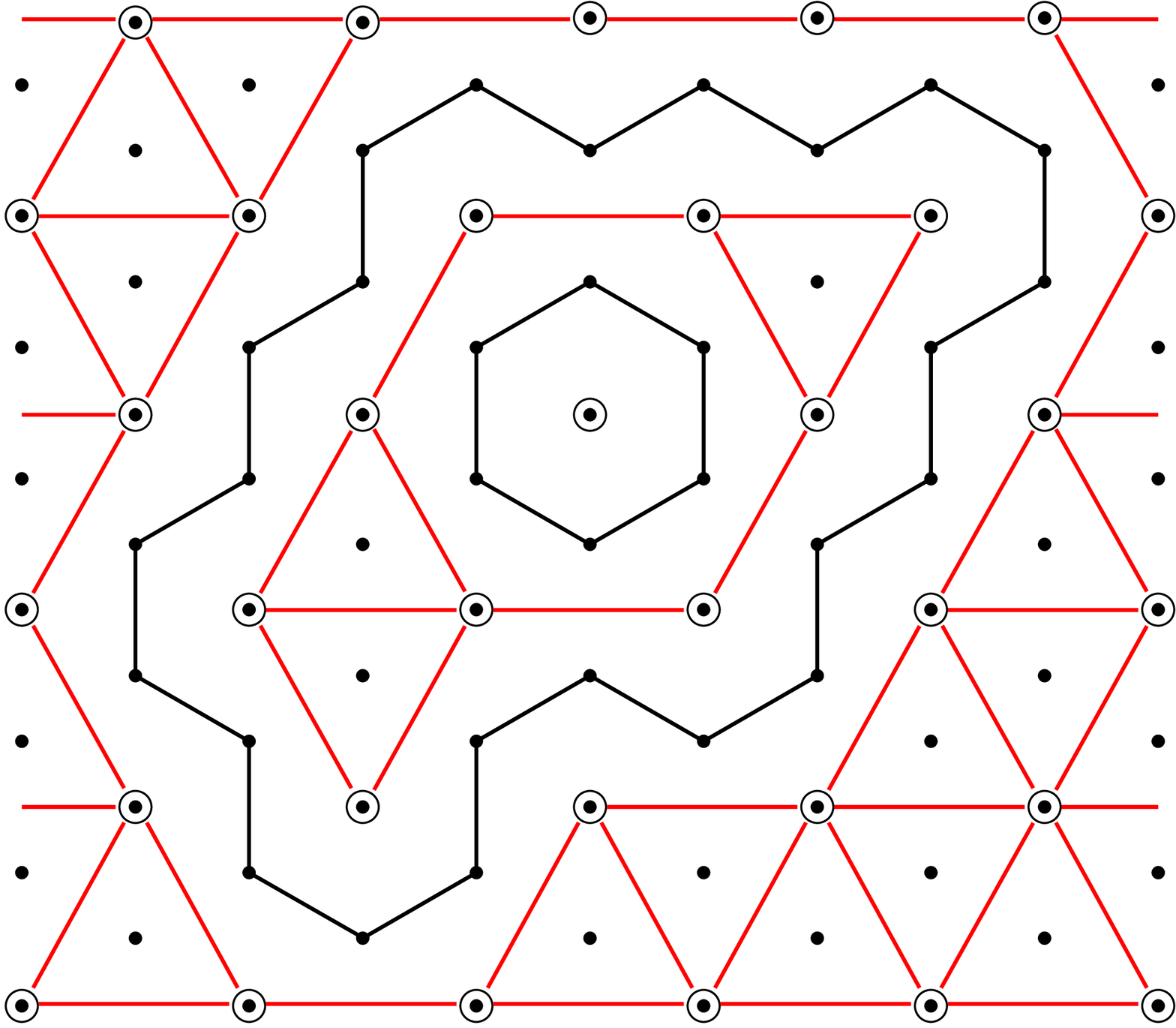}
\end{center}
\caption{Equivalence of the centered triangular model with the O($n$) loop 
model on the honeycomb lattice, under the condition that triangles with three
different Potts spins are excluded. This condition applies in the limits
$J \to \pm \infty$ when $K>>-2|J|$. Triangular edges between equal Potts
spins are shown in red. The honeycomb loops separate unequal triangular spins.
Centered spins connected by a black loop segment are equal.
}
\label{mapping}
\end{figure}

The construction of an O(2) loop configuration from an allowed $q=3$
Potts spin configuration is formulated as follows. We first note that
elementary triangles with three different spins on the triangular
vertices would cost an energy $\propto |J|$, and are therefore excluded.
Each allowed triangle contains precisely one or three edges connecting
equal spins. This is illustrated in Fig.~\ref{mapping} by erasing all
triangular edges connecting unequal spins, leading to a graph with
one or three edges remaining about each elementary triangular face.
Next, construct a dual graph from edges connecting each pair of dual
sites if not separated by a remaining triangular edge. Thus, each dual
site connects to zero or two edges on the dual honeycomb lattice.
In this way one obtains a configuration of closed loops on the
honeycomb lattice.
Thus the triangular neighbor spins are equal if and only if they are not
separated by such a loop. The introduction of a new loop in a region of
triangular Potts spins equal to $\sigma_{\rm old}$ will thus change the
inside spin configuration. The spin degrees of freedom allow the centered
spins on the loop to take two values 
\begin{equation}
\tau =\sigma_{\rm old}  \pm 1 \mod 3
\label{newtau}
\end{equation}
(with the convention $1 \leq k \mod 3 \leq 3$), while $\sigma_{\rm old}$
remains the value of the triangular spins directly outside the loop. 
Then, each triangular spin along the inside perimeter of the loop must
change its old value $\sigma_{\rm old}$ in
\begin{equation}
\sigma_{\rm new}=\sigma_{\rm old} \mp 1 \mod 3\, ,
\label{newsp}
\end{equation}
so that these inside spins are unequal to the spins on the loop,
and unequal to the outside spins.  Application of this rule (\ref{newsp})
to {\it all} spins inside the new loop guarantees that the energy
changes are restricted to the bonds crossing that loop,
even if the region inside the loop contains further loops.
The spin degeneracy expressed in Eqs.~(\ref{newtau}) and (\ref{newsp})
translates into a weight factor 2 for each loop on the honeycomb lattice.

To complete the mapping onto the O($n$) model, we still have to obtain
the weight $x$ of each loop segment. This is done by comparing the 
weight of a loop to the weight ratio of spin configurations with and
without a loop. In the O($n$) model, the weight
of a loop consisting of $n_{\rm s}$ loop segments is
$w_{\rm loop}=n x^{n_{\rm s}}$, while the vacuum has weight
$w_{\rm vac}=1$.
The loop intersects $n_{\rm s}$ triangles with weight $W_{\rm 112}$.
Removal of this loop changes their weight into $W_{\rm 111}$. Thus the
weight of the loop is $2 [W_{\rm 112}/W_{\rm 111}]^{n_{\rm s}}$ in the spin
language. %, where the prefactor 2 reflects the spin degeneracy
%of the centered (honeycomb) sites located on the loop. 
The expression for the weight ratio depends on the sign of $J$.
\begin{enumerate}
\item
In the case $J\to -\infty$ the terms in $W_{\rm 111}$ and $W_{\rm 112}$ that
contain $J$ vanish, and $W_{\rm 112}/W_{\rm 111}=\exp (-K)/2$.
\item
For $J\to +\infty$ the terms in $W_{\rm 111}$ and $W_{\rm 112}$ with the
largest prefactors of $J$ survive, and $W_{\rm 112}/W_{\rm 111}=\exp(-K-J)$.
\end{enumerate}
A comparison of the weights of the configurations with and without a
loop in both representations directly determines the O($n$) loop weight 
$n$ and the relation between $x$ and the Potts couplings.
\begin{equation}
w_{\rm loop}/w_{\rm no\,loop}=n x^{n_{\rm s}}=2 [\exp(-K)/2]^{n_{\rm s}}
~{\rm for}~ J \to -\infty \,,
\end{equation}
\begin{equation}
w_{\rm loop}/w_{\rm no\,loop}=n x^{n_{\rm s}}=2 [\exp(-K-J)]^{n_{\rm s}}
~{\rm for}~ J \to +\infty \,,
\end{equation}
The partition sum of the loop model is defined as 
\begin{equation}
Z_{\rm loop}(x, n)=\sum_{\mathcal G} x^{n_{\rm b}} n^{n_{\rm l}}\, ,
\end{equation}
where the sum is on all loop configurations ${\mathcal G}$, $n_{\rm b}$
is the number of honeycomb edges covered by ${\mathcal G}$, and $n_{\rm l}$
is the number of loops.
The prefactor in its relation with the partition sum $Z_{\rm ctri}$ of 
the spin model can, for instance, be found from a comparison between the
Boltzmann factors of the loop vacuum in the two representations. 
The resulting relation between the two models is summarized as
%\newpage
\begin{equation}
Z_{\rm ctri}(K,J)=2^{2N} e^{3NK}Z_{\rm loop}(x,n)\, ,~~~n=2,~~~x= e^{-K}/2 
~{\rm for}~ J \to -\infty \,,
\label{Kx}
\end{equation}
\begin{equation}
Z_{\rm ctri}(K,J)=e^{3N(K+2J)}Z_{\rm loop}(x,n)\, ,~~~n=2,~~~x= e^{-K-J} 
~{\rm for}~ J \to +\infty \,,
\label{KJx}
\end{equation}
where $N$ is the number of triangular sites. The free energies, per 
triangular and honeycomb site respectively, are thus related as
$f_{\rm ctri}(K)=2 \ln 2+3K+2f_{\rm loop}(x,n)$ for $J \to -\infty $
and as $f_{\rm ctri}(K)=3K+6J+2f_{\rm loop}(x,n)$ for $J \to +\infty $.

\subsection{The fully packed loop model}
For $K \to -\infty$, but still subject to Eq.~(\ref{Jinf}), the weight
of the honeycomb edges not covered by a loop vanishes, and we obtain
the fully packed O(2) model. This model displays a rather special
behavior \cite{BNfpl,KH}, for instance, its conformal anomaly was found
to be equal to 2. This value can be interpreted in terms of two SOS-like
degrees of freedom, one of which comes from the O(2) model, and the other
from the equivalence \cite{BNfpl} of the fully packed loop model with the
triangular SOS model \cite{NHB}.
Using the O(2) loop representation, we have extended the transfer-matrix
calculations of the free energy up to finite-size $L=21$. 
The conformal anomaly can be estimated for each single system size that is
a multiple of 3, using the free energy per honeycomb site for the infinite
system, which is known from an exact result by Baxter \cite{Bax3col} as
\begin{equation}
\lim_{x \to \infty}f_{\rm loop}(x)-x=\frac{1}{2} \ln \prod_{i=1}^{\infty}
\frac{(3i-1)^2}{3i(3i-2)} \, ,
\end{equation}
which can be approximated as
$\lim_{x \to \infty}f_{\rm loop}-x=0.189560048316\cdots$.
Taking into account the geometric factor $\zeta=2/\sqrt 3$, which is needed 
to obtain the free energy density of the honeycomb lattice instead of the
free energy per site, the finite-size estimates are \cite{BCN,Affl}
\begin{equation}
c_{\rm est}(x,L)=4\sqrt 3 L^2[f_{\rm loop}(x,L)-f_{\rm loop}(x,\infty)] /\pi
\label{ccal}
\end{equation}
The usual extrapolation of these estimates by power-law fits, assuming
power-law corrections as $L^{-2}$, yields iterated estimates of $c(\infty)$
close to 2, with differences of a few times $10^{-2}$, suggesting the
presence of a logarithmic correction. Including an extrapolation step as
$c_{\rm est}(x \to \infty,L) \simeq c(x \to \infty) 
\left[ 1+ \frac{a}{L^2(b+\ln L)} \right]$ led to a better apparent
convergence, with the last two iteration steps within $10^{-4}$ from
$c=2$.
\subsection{Phase changes induced by $K$}
\begin{figure}[bthp]
\begin{center}
\leavevmode
\includegraphics[width=12.cm,angle=0]{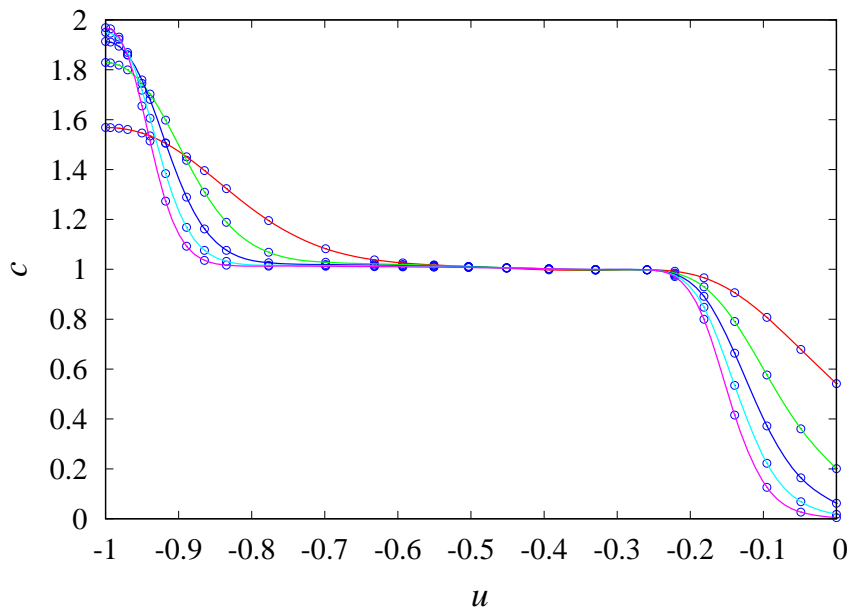}
\end{center}
\caption{Finite-size estimates of the conformal anomaly $c$ as a 
function of $u=e^K-1$ of the $J \to -\infty$ centered triangular $q=3$
antiferromagnetic Potts model. Estimates for system sizes $L$ that are
multiples of 3 are obtained by solving for $c$ in Eq.~(\ref{ccal}),
from the free energies for systems  with sizes $L$ and $L-3$. 
Results are shown for $L=6$, 9, 12, 15, and 18. Larger system sizes
correspond with steeper curves.
This behavior applies as well in the limit $J \to +\infty$ after
redefining $u=e^{K+J}-1$. These results still depend on the absence
of type (1,2,3) triangles, implying $K/|J|>-2$.
}
\label{cest}
\end{figure}

For finite values of $|K|$, but still subject to condition (\ref{Jinf}),
the model is still exactly equivalent with the O(2) loop model, but no
longer fully packed.
The fugacity of empty honeycomb vertices is relevant \cite{BNfpl}, and
crossover takes place to the universal behavior of the dense phase
of the O(2) model which has $c=1$. This crossover is illustrated by the
finite-size estimates of the conformal anomaly in Fig.~\ref{cest}.
This figure uses the parametrization $u=e^K-1$ so that the whole
antiferromagnetic range $K<0$ can be included.

When $|K|$ is sufficiently lowered, the spin model undergoes an
infinite-order Berezinskii-Kosterlitz-Thouless transition \cite{BKT}
to a state with ferromagnetic order on the triangular sites,
and disordered spins on the honeycomb sites. The transition to the
$c=0$ phase (loops diluted, and ferromagnetic in the language of 
the spins on the triangular lattice) is visible in the right-hand side
of Fig.~\ref{cest}, and was numerically located from the requirement
\begin{equation}
X_h (K,L)\equiv L/[2 \pi \xi(K,L)]=2/9,
\end{equation}
where 2/9 is the expected value of the magnetic dimension of the transition;
see, for instance the similar analysis in Ref.~\cite{triigs}.
We thus estimate $K_{\rm BKT}=-0.3465 (1)$, in a good agreement with the
exact value $-\ln(2)/2$ which follows from Eq.~(\ref{Kx}) and
$x_{\rm c}=1/\sqrt 2$ \cite{N82}.

\begin{figure}[bthp]
%\begin{center}
\leavevmode
\includegraphics[scale=0.8,width=12.cm,angle=0]{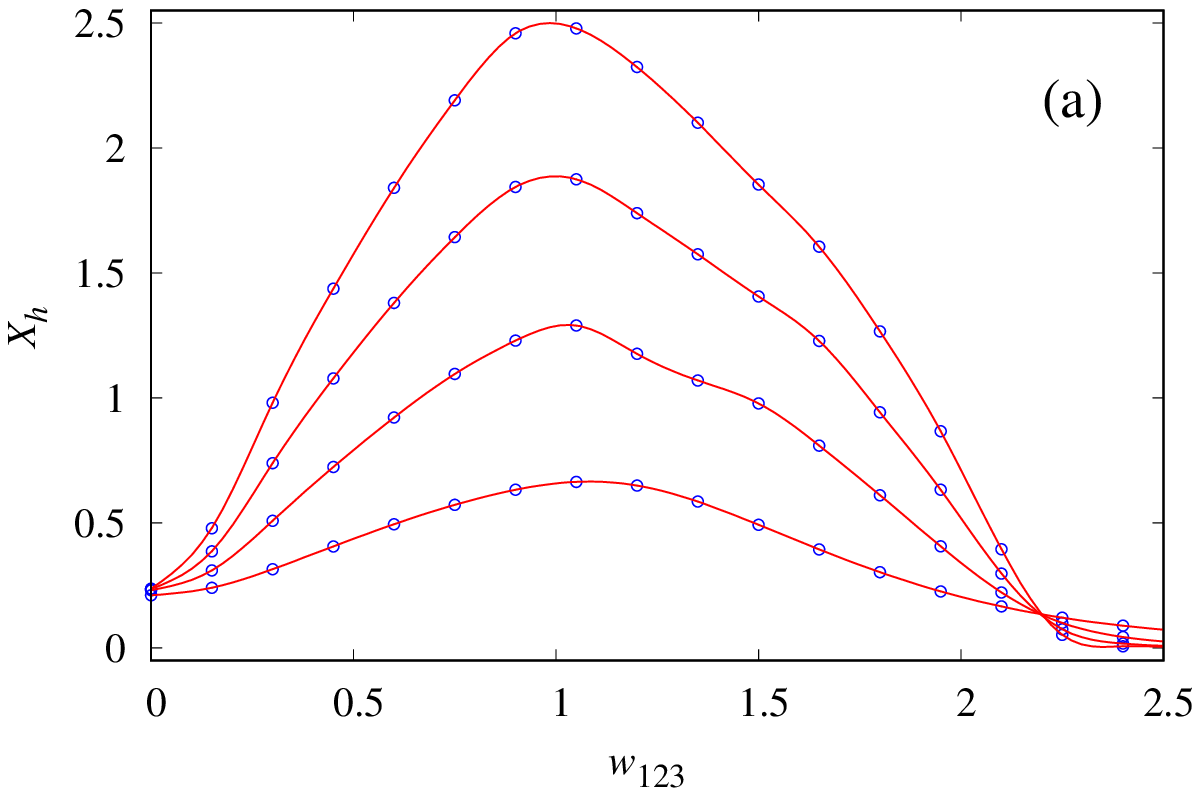}
\vspace{-1mm} \\ \hspace*{-10mm}
\includegraphics[scale=0.66,trim=0 3mm 0 -3mm]{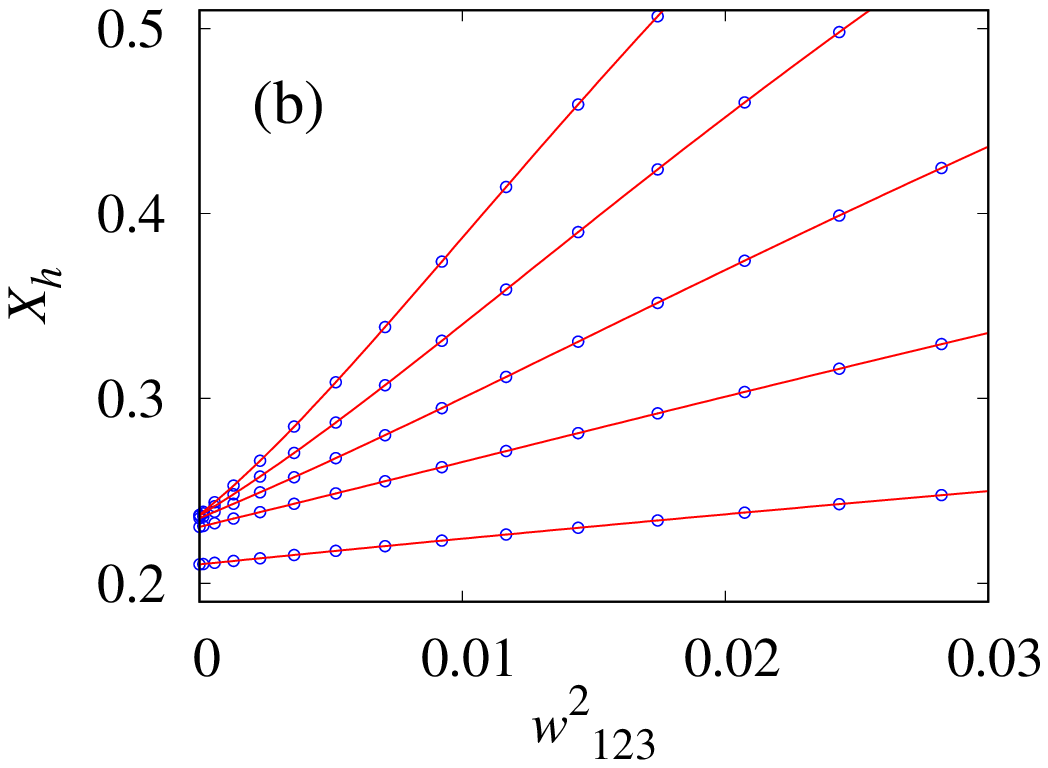}
\hspace*{-10mm} %\vskip -1mm
\includegraphics[scale=0.62,trim=0 -2mm 0 2mm]{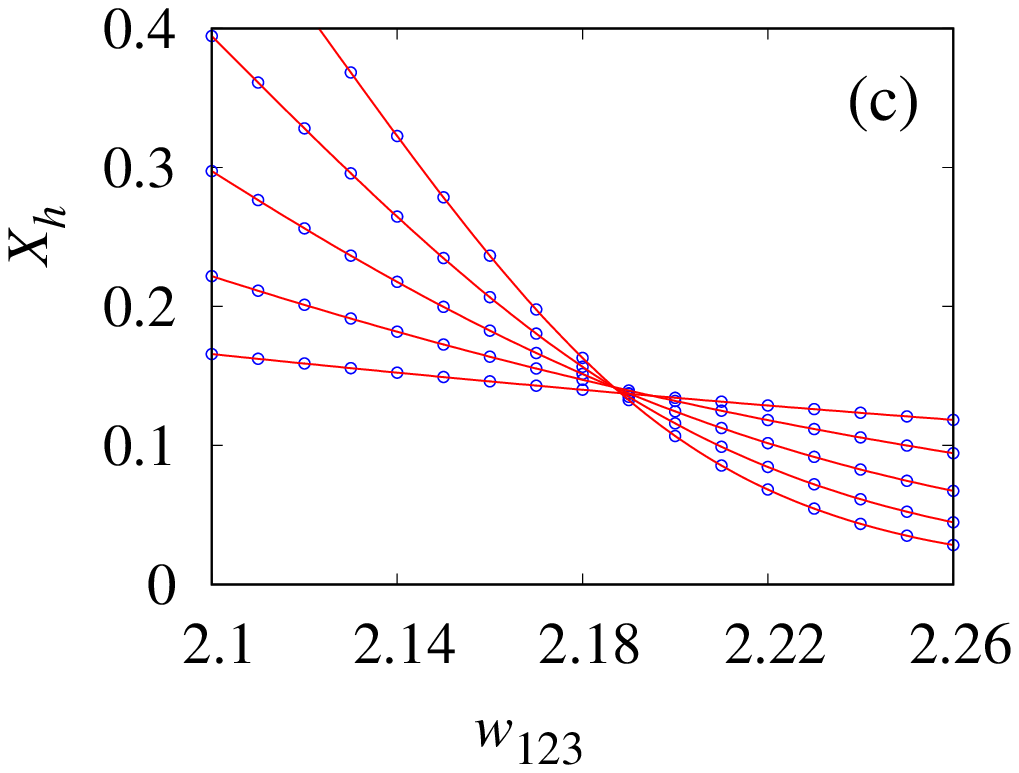} 
%\end{center}
\caption{Finite-size results for the scaled magnetic gaps near the 
transition to the (1,2,3) phase, versus the rescaled weight $w_{\rm 123}$,
in the limit $|J| \to \infty$.
Figure (a) shows the whole range of interest, for system sizes 
$L$=3, 6, 9, and 12.  Curves are shown connecting discrete data points.
Larger system sizes correspond with steeper curves.
More detailed pictures of the data on the left and right parts are shown
in Figs. (b) and (c) for finite system sizes $L$=3, 6, 9, 12, and 15.
}
\label{xh123}
\end{figure}

The mapping on the O($n$) model relies on the condition (\ref{Jinf}).
Next, we drop the condition that limits $K$ in Eq.~(\ref{Jinf}), while
maintaining the limit $|J| \to \infty$. Then, type (1,2,3) triangles are no
longer excluded, and the mapping on the O($n$) model is no longer valid.
One expects a transition near $K \approx 2J$ to the antiferromagnetic phase.
We investigated this point using transfer-matrix calculations, based on
finite-size scaling of the magnetic correlation length.
The behavior of the scaled gaps, defined as
$X_h(L)\equiv L/[2\pi \xi(K,L)]$, is displayed in Figs.~\ref{xh123}
in the vicinity  of the transition, versus the rescaled weight $w_{\rm 123}$.
The rescaled weights are obtained by dividing out $W_{\rm 112}$,
{\it i.e.}, $w_{\rm 123}\equiv W_{\rm 123}/W_{\rm 112}$, $w_{\rm 112}=1$
and $w_{\rm 111}=0$.
The apparent divergence of $X_h(L)$ with $L$, shown in Fig. \ref{xh123}(a)
indicates the existence of an intermediate disordered phase for finite 
$2|J|+K$. Thus the disordered phase extends all the way to $T=0$. 
The behavior on the left-hand side, highlighted in Fig. \ref{xh123}(b),
illustrates that the critical state is destroyed by a nonzero $w_{\rm 123}$.
Figure (b) uses $w^2_{\rm 123}$ on the horizontal scale, because (1,2,3)
triangles appear in pairs. One thus expects a finite-size dependence
according to
\begin{equation}
X_h [w_{\rm 123},L]= X_h + \sum_k p_k w^{2k}_{\rm 123}  L^{k y_w} +\ldots \, ,
\end{equation}
where $y_w$ is the renormalization exponent describing the fugacity of
a pair of (1,2,3) triangles.
Numerical fits to the transfer-matrix data lead to $y_w \approx 1.4$,
with poor apparent convergence. This result seems consistent with
$y_w =3/2$, as expected on the basis of the O($n$) magnetic dimension
$X_h^{\rm FPL}=1/2$ reported in Ref.~\cite{BNfpl} (which is different
from the present Potts dimension $X_h$). The relation with the O($n$)
dimension follows from the fact that a type (1,2,3) triangle corresponds,
along the lines of the mapping described in Sec.~\ref{mapon}, 
with the open end of an O($n$) loop segment.

The right hand side of Fig. \ref{xh123}(a), and the enlarged version
in Fig. \ref{xh123}(c), display intersections
associated with the transition to the antiferromagnetic phase.
Numerical analysis of the intersection points locates this transition
near $K-2J=0.631$. 
The data do not permit a clear answer about the type of transition, but
are suggestive of a weak first-order transition. The amplitude
of the correlation length, as determined for finite sizes up to $L=18$,
could not be reliably extrapolated, but might seem to correspond with
a magnetic dimension of about $X_h=0.14$.

Thus far we have considered the antiferromagnetic limit $J \to -\infty$,
but similar phenomena are also be expected for  $J \to +\infty$.
For $K>>- 2J$ the (1,2,3) triangles are then excluded, and the  mapping
on the O($n$) model applies.
Following the same line of reasoning as for the antiferromagnetic case, 
one finds  from Eq.~(\ref{KJx}) that an infinite-order transition to the
ferromagnetic phase occurs at $K+J=\ln(2)/2$.
Finally, the transition to the antiferromagnetic phase takes place 
close to $K+2J=0.631$, mirroring the transition for $J \to -\infty$.
The location is verified by Monte Carlo calculations.

\section{Conclusion}
Our investigation of the phase diagram of the $q=3$ Potts model on
the centered triangular lattice in the $(K,J)$ plane shows the existence
of three phases: a ferromagnetic phase dominated by  one of the three
Potts states; an antiferromagnetic phase where the three different
Potts states condense on different triangular sublattices; and an
intermediate  disordered phase dominated by triangles containing two
different Potts states.

In the limits $J \to \pm \infty$, the disordered phase evolves into a 
state with partial order. There exist, in these limits, infinite
ranges of $K$ where the model is critical, and where it is equivalent
with the fully packed O(2) loop model on the honeycomb lattice. In
addition there are ranges of $K$ at the ferromagnetic sides, where the
mapping on the O(2) loop model is still valid, but where it is no longer
fully packed. The ferromagnetic transitions are of infinite order in
these limits. The situation reminds of the triangular Ising model in a
field, which also undergoes a three-state Potts transition, changing
into an infinite-order transition when $T \to 0$ \cite{NHB,QWB}.

On the antiferromagnetic side of the critical ranges, there are 
ranges of $K$ where the critical state is destroyed by the nonzero
weight of triangles with three different Potts spins. While these
disordered ranges are, strictly speaking, infinitely wide on the scale
of $K$, they are restricted to $K/|J|=-2$ when $|J| \to \infty$. 
The transitions between the disordered phase and the antiferromagnetic
phase are probably discontinuous for all $J$.

\acknowledgments
Z.~F. would like to thank C.~X. Ding and W.~G. would like to 
thank F.~Y. Wu for valuable discussions. This research was
supported by the National Natural Science Foundation of China under
Grants No. 11775021, No. 11734002, and No. 11447154, by the Ninth Group of Key
Disciplines in Henan Province under Grant No. 2018119, and by the
Natural Science Foundation of the Henan Department of Education under
Grant No. 18B430012.  H.~B.  acknowledges hospitality extended to him
by the Faculty of Physics of the Beijing Normal University.

%\newpage


\begin{thebibliography}{widest-label}
\bibitem{Potts}
R. B. Potts, Proc. Camb. Phys. Soc. {\bf 48}, 106 (1952).
\bibitem{esmsm}
R.~J. Baxter, {\em Exactly Solved Models in Statistical Mechanics}
(Academic Press, London-New York, 1982).
\bibitem{CFT}
P.~Di Francesco, P. Mathieu, and D. S\'en\'echal,
{\em Conformal Field Theory} (Springer-Verlag, New York, 1997).
\bibitem{Wurev}
F.~Y. Wu, Rev. Mod. Phys. {\bf 54}, 235 (1982).
\bibitem{Adler} J. Adler, A. Brandt, W. Janke, and S. Shmulyian,
J. Phys. A {\bf 28}, 5117 (1995).
\bibitem{dicedRK}
R.~Koteck\'y, J.~Salas and A.~D.~Sokal, Phys. Rev. Lett. {\bf 101},
030601 (2008).
\bibitem{honeycombJS}
J.~Salas, J. Phys. A {\bf 31}, 5969 (1998).
\bibitem{BH}
J.~K. Burton Jr. and C.~L. Henley, J. Phys. A: Math. Gen. {\bf 30}, 8385 (1997).
\bibitem{Kolafa}
J. Kolafa,  J. Phys. A: Math. Gen. {\bf 17}, L777 (1984).
\bibitem{denNijs}
M.~P.~M. den Nijs, M.~P. Nightingale, and M. Schick,
Phys. Rev. B {\bf 26}, 2490 (1982).
\bibitem{SS}
J. ~Salas and A.~D. Sokal, J. Stat. Phys. {\bf 92}, 729 (1998).
\bibitem{jplv}
J.~P. Lv, Y. Deng, J.~L. Jacobsen, and J. Salas, J. Phys. A: Math.
Theor. {\bf 51}, 365001 (2018).
\bibitem{KF}
P.~W. Kasteleyn and C.~M. Fortuin, J. Phys. Soc. Jpn. {\bf 26}
(Suppl.), 11 (1969);
C.~M. Fortuin and P.~W. Kasteleyn, Physica (Amsterdam) {\bf 57}, 536 (1972).
\bibitem{WG} F.~Y. Wu and W.~A. Guo, Phys. Rev. E {\bf 86}, 020101 (2012).
\bibitem{Huang}
Y. Huang, K. Chen, Y. Deng, J.~L. Jacobsen, R. Koteck\'y, J. Salas,
A.~D. Sokal, and J.~M. Swart, Phys. Rev. E {\bf 87}, 012136 (2013).
\bibitem{kotecky1}
R. Koteck\'y, Phys. Rev. B {\bf 31}, 3088 (1985).
\bibitem{Chen}
Q.~N. Chen, M.~P. Qin, J. Chen, Z.~C. Wei, H.~H. Zhao, B. Normand,
and T. Xiang, Phys. Rev. Lett. {\bf 107}, 165701 (2011).
\bibitem{Deng}
Y. Deng, Y. Huang, J.~L. Jacobsen, J. Salas, and A.~D. Sokal,
Phys. Rev. Lett. {\bf 10}, (2011).
\bibitem{Syozi}
I. Syozi, in {\it Phase Transitions and Critical phenomena}, edited
by C. Domb and M.~S. Green (Academic, London, 1972), Vol. 1.
\bibitem{Cardyxi} J.~L. Cardy, J. Phys. A {\bf 17}, L385 (1984).
\bibitem{CG}
B. Nienhuis, in {\it Phase Transitions and Critical Phenomena}, Vol. 11,
eds. C. Domb and J.~L. Lebowitz (Academic, London, 1987).
\bibitem{FSS}
For reviews, see e.g. M.P. Nightingale in {\it Finite-Size Scaling and
Numerical Simulation of Statistical Systems}, edited by V. Privman
(World Scientific, Singapore 1990); and M.~N. Barber in {\it
 Phase Transitions and Critical Phenomena}, edited by C. Domb
and J.L. Lebowitz (Academic, New York 1983), Vol. {\bf 8}.
\bibitem{Baxtrcp}
R.~J. Baxter, H.~N.~V. Temperley and S.~E. Ashley,
Proc. Roy. Soc. London, Ser. A {\bf 358}, 535 (1978).
\bibitem{JS}
J.~L. Jacobsen and C.~R. Scullard, J. Phys. A {\bf 46}, 075001 (2013).
\bibitem{Wea} M.~X. Wang, J.~W. Cai, Z.~Y. Xie, Q.~N. Chen, H.~H. Zhao, and
Z.~C. Wei, Chin. Phys. Lett. {\bf 27}, 076402 (2010).
\bibitem{N82}  B. Nienhuis, Phys. Rev. Lett. {\bf 49}, 1062 (1982).
\bibitem{BNfpl} H.~W.~J. Bl\"ote and B. Nienhuis,
Phys. Rev. Lett. {\bf 72}, 1372 (1994).
\bibitem{KH} J. Kondev and C.~L. Henley, Phys. Rev. Lett. {\bf 73}, 2786 (1994).
\bibitem{NHB}
B. Nienhuis, H.~J. Hilhorst and H.~W.~J. Bl\"{o}te,
J. Phys. A {\bf 17}, 3559 (1984).
\bibitem{Bax3col} R.~J. Baxter, J. Math. Phys. {\bf 11}, 784 (1970).
\bibitem{BCN}
H.~W.~J. Bl\"{o}te, J.~L. Cardy and M.~P. Nightingale,
Phys. Rev. Lett. {\bf 56}, 742 (1986).
\bibitem{Affl}
I. Affleck, Phys. Rev. Lett. {\bf 56}, 746 (1986).
\bibitem{BKT}
V.~L. Berezinskii, Zh. Eksp. Teor. Fiz. {\bf 59}, 907 (1970) [Sov. Phys. JETP
{\bf 32}, 493 (1971)];
J.~M. Kosterlitz and D.~J. Thouless, J. Phys. C {\bf 5}, L124 (1972);
J.~M. Kosterlitz and D.~J. Thouless, J. Phys. C {\bf 6}, 1181 (1973).
\bibitem{triigs}
H.~W.~J. Bl\"ote and M.~P. Nightingale, Phys. Rev. B {\bf 47}, 15046 (1993).
\bibitem{QWB}
X.~F. Qian, M. Wegewijs and H.~W.~J. Bl\"ote,
Phys. Rev. E {\bf 69}, 036127 (2004).
\end{thebibliography}
\end{document}